\documentclass{article}
\usepackage{amssymb}
\usepackage[dvips]{graphicx}
\begin{document}
\begin{flushright}

{\bf \large SLIM 2005/5}\\
\vspace{2mm}
 Bologna, \today \\
\end{flushright}

\begin{center}
{\large {\bf SEARCH FOR NUCLEARITES\\ WITH THE SLIM DETECTOR}}

\vspace{5mm}
S. Balestra, S. Cecchini$^1$, G. Giacomelli,
M. Giorgini, A. Kumar$^2$ G. Mandrioli,
S. Manzoor$^3$, E. Medinaceli$^4$, L. Patrizii,
V. Popa$^5$,  V. Togo

\vspace{3mm}
{\it Dipartimento di Fisica dell'Universit\`{a} and INFN Sezione di Bologna,
 I-40127 Bologna,
Italy\\
1. also IASF/INAF, I-40129 Bologna, Italy\\
2. also Dept. of Physics, Sant Longowal Inst. of Eng. \& Tech., Longowal, 148106 India \\
3. also PINSTECH, P.O. Nilore Islamabad, Pakistan\\
4. also Inst. Invest. Fisicas, La Paz, Bolivia\\
5. also ISS, R-77125 Bucharest, Romania\\
}
\end{center}

\vspace{5mm}

{\bf Abstract}  We discuss the properties of
cosmic ray nuclearites, from the point of view of their search with large
nuclear track detector arrays exposed at different
altitudes, in particular with the SLIM experiment
 at the Chacaltaya high altitude lab (5290 m a.s.l.).
We present calculations concerning their propagation
in the Earth atmosphere and discuss
their possible detection  with CR39 and Makrofol nuclear track detectors.

\section{Introduction}

Strange Quark Matter (SQM) could be the ground state of
quantum chromodinamics \cite{witten}. The initial hypothesis assumed that SQM is
made of
$u$, $d$ and $s$ quarks in nearly equal proportions (with some electron component
in weak equilibrium). Recently it was shown that the so called
``color-flavor locked" (CFL) SQM, characterized by a Cooper-like pairing between
different quarks, could be even more stable \cite{madsen}.

 SQM is expected to have a density
slightly larger than ordinary nuclear matter \cite{witten,madsen}; the
relation between the mass $M$ of SQM lumps and their baryonic number $A$
( $\simeq$ one third
of the number of constituent quarks) would be
\begin{equation}
M(\mbox{GeV}) \lesssim 0.93A.
\label{masa}
\end{equation}

It was hypothesized
that ``nuggets" of SQM, with masses from those of heavy nuclei
to macroscopic values, produced in the Early Universe or
in violent astrophysical processes, could be present in the cosmic radiation
(the so-called {\em nuclearites}) \cite{ruhula} \footnote{The attention of the authors
was mostly focused on relatively large mass nuggets of SQM, so that the microscopic
properties of the nuclearites would not be relevant. This is suggested also by
the name they proposed for the newly postulated objects, ``nuclearites": a
combination between ``nuclei" and ``meteorites". Note that a nuclearite is an
electrically neutral state composed of a SQM ``nucleus" and electrons.}.
 SQM should have a relatively small positive electric charge, eventually
neutralized by an electron cloud. If the size of the SQM is large
(corresponding to masses $M \gtrsim 10^7$ GeV),
some of the electrons could be in chemical
 equilibrium inside the quark core \cite{kasuya}. Nuclearites larger than 1 \AA,
 ($M \geq 8.4 \times 10^{14}$GeV) would contain all electrons inside the quark core
and thus would be completely neutral \cite{ruhula}.
 A qualitative picture of
nuclearites may be found in \cite{noi}.

An upper limit for the flux of nuclearites may be obtained assuming
that they represent the main contribution to the local Dark Matter (DM) density,
$\rho_{DM} \simeq 10^{-24}$ g cm$^{-3}$ \cite{ruhula},
\begin{equation}
\Phi_{max} = \frac{\rho_{DM} v}{2 \pi M},
\label{dm}
\end{equation}
where $v$ and $M$ are the nuclearite average velocity and mass, respectively.

The aim of this note is to discuss the possibility to detect nuclearites using
large area Nuclear Track Detectors (NTDs)
 at mountain altitude (in particular with the SLIM
experiment \cite{slim,slimnou,slimsimainou}).
 We classify nuclearites in three different mass ranges (and sizes).
\begin{itemize}
\item {\bf Low mass nuclearites (LMNs), or {\em strangelets}},
with masses between those of ordinary nuclei $(A \lesssim 300)$
and a multi-TeV mass. The upper bound
is an ad-hoc one; an indirect definition of this category could be that many of
the properties of LMNs would contradict the assumptions made in \cite{ruhula}
 summarized in Section 3.
We shall discuss the case of LMNs in Section 2.
\item {\bf Intermediate mass nuclearites (IMNs) or, simply {\em nuclearites}},
with
masses large enough to be well described by the hypothesis made in \cite{ruhula},
but smaller than about $10^{22}$ GeV (for $M>10^{22}$ GeV nuclearites
would traverse the entire Earth). The mass lower limit of IMNs may be
 about $10^8$ GeV, above
which nuclearites could be detected by experiments performed in the upper
atmosphere (see also Fig.
\ref{conc}). Assuming that they would travel in space with typical galactic
velocities ($\beta = v/c \simeq 10^{-3}$), they would be stopped by the Earth,
so they would reach detectors only from above.
The main properties of IMNs are summarized in Section 3.
\item{\bf ``Macroscopic" nuclearites}, with
masses $M > 10^{22}$ GeV; assuming galactic velocities, such nuclearites would traverse
the Earth. They differ from IMN's only by size.
We shall not discuss this case, since the expected sensitivity of
SLIM (and of similar experiments) will not compete with the limit
obtained by MACRO in this mass range \cite{macroini,macro,macrogg}.

Much heavier nuclearites ($ M \gtrsim 10^{28}$ GeV)
 could be
observed as abnormal sismic events \cite{seism,seism2} \footnote{In ref.
\cite{seism} there was a claim of observing a candidate, but it was discarded
in \cite{seism2} because of timing uncertainities.}.
\end{itemize}

Calculations describing the production (through binary strange stars tidal disruption)
and the galactic propagation of cosmic ray nuclearites were recently published
\cite{mad}. The results could be valid  as
orders of magnitude for the entire mass
range of interest; we shall use the predicted fluxes (at the Earth level) as reference
values.

Searches for nuclearites (mostly IMNs) were performed by different experiments
\cite{naka,orito}. The  best flux upper limit was set by the MACRO
experiment:  for nuclearites
with $\beta \simeq 10^{-3}$,
the 90\% C.L. upper limit is
at the level of $2 \times 10^{-16}$ cm$^{-2}$sr$^{-1}$s$^{-1}$ in the mass range
$10^{14}$ GeV $<M< 10^{22}$ GeV \cite{macroini,macro,macrogg}\footnote{
This is twice the flux limit obtained for relativistic
GUT magnetic monopoles \cite{macromono}, as it refers only to down-going nuclearites.}.

SLIM is a large area experiment (440 m$^2$) installed at the Chacaltaya high altitude
laboratory since 2001; an additional 100 m$^2$ were installed at Koksil,
Pakistan, since 2003~\footnote{
The calculations presented in this report refer to the Chacaltaya location only.}.
 With an
average exposure time of about 4 years, SLIM would be sensitive to a flux of
downgoing exotic particles (magnetic monopoles, nuclearites and Q-balls)
 at a level of  $10^{-15}$ cm$^{-2}$sr$^{-1}$s$^{-1}$.

\section{Low mass nuclearites (LMNs)}

SQM  should be
stable for all masses larger than about 300 GeV  \cite{ruhula}.
Nuclearites
with masses up to the TeV region
 could be ionized and could
 be accelerated  to relativistic
velocities by the same astrophysical mechanisms of normal nuclei of the
primary cosmic radiation (CR).

LMNs would interact with detectors (in particular NTDs) in ways similar to
heavy ions, but with different $Z/A$. There are different calculations in relative
agreement with a possible candidate with $M \simeq 370$ GeV and a charge
$Z \simeq 14$ \cite{cand}.

\begin{figure}
\begin{center}
\vspace{-25mm}
\includegraphics[angle=0, width=0.8\textwidth]{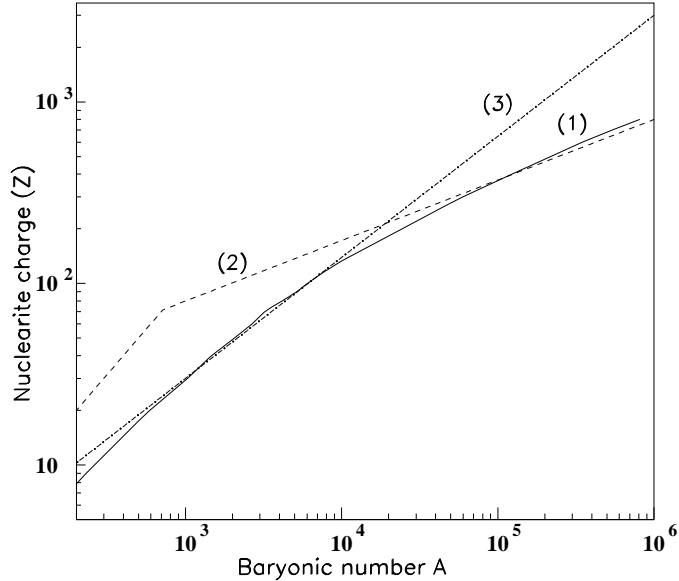}
\caption{Low mass nuclearites (LMN) charge versus mass for different hypotheses
discussed in refs. \cite{kasuya,heis,madsen}. See text for
details.}
\label{a-z}
\end{center}
\end{figure}

In ref. \cite{kasuya} SQM is described in analogy
with the liquid-drop model of normal nuclei; the obtained charge versus
mass relation is shown in Fig. \ref{a-z} by the solid line, labeled ``(1)".
Other
authors found different relations:  $Z \simeq 0.1 A$
for $A \lesssim 700$  and $Z \simeq 8 A^{1/3}$ for larger
masses \cite{heis}: this charge to mass relation is shown in Fig. \ref{a-z}
as the dashed line, labeled ``(2)".
In \cite{madsen} it was assumed that quarks with different
color and flavor quantum numbers form Cooper pairs inside the SQM (the so-called
color-flavor locked phase), increasing the stability of the strangelets.
In this case, the charge relation would be
$Z \simeq 0.3 A^{2/3}$, shown as the dash-dotted line in Fig. \ref{a-z} labeled
``(3)".

\begin{figure}
\begin{center}
\vspace{-40mm}
\includegraphics[angle=0, width=0.8\textwidth]{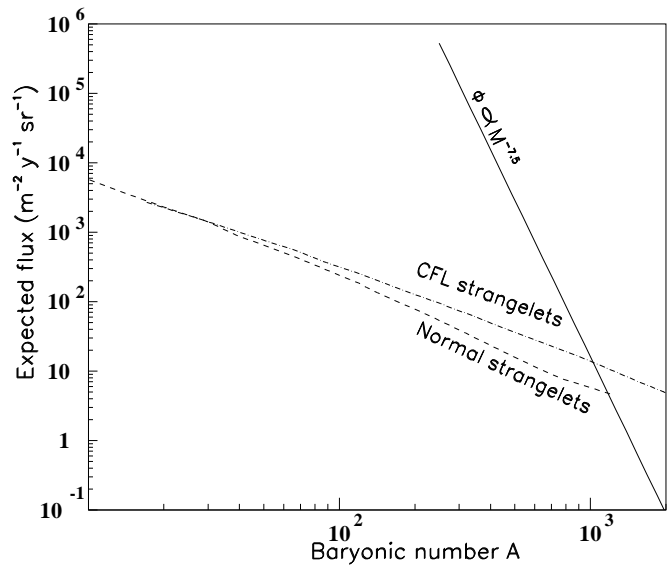}
\caption{Expected fluxes for LM strangelets in the CR near the Earth.
The solid line corresponds to the assumption that their abundances
follow the same rule as heavy CR nuclei \cite{wilk1}. The dashed and dot-dashed
lines are from Ref. \cite{mad}, and refer to ``normal" and CFL strangelets.}
\label{flux}
\end{center}
\end{figure}

Several CR experiments reported candidate events that would suggest anomalously
low charge to mass (Z/A) ratios, which could correspond to those
expected for SQM \cite{kasuya}. Such candidates are
reviewed in \cite{wilk1,raha}. As strangelets with masses not much higher than those of
ordinary nuclei could have the same origin as CR heavy nuclei, their
abundances in
the cosmic radiation could follow the same mass dependence, $\Phi
\propto M^{-7.5}$,  \cite{wilk1}. The existing candidates do
not contradict such an hypothesis.
The solid line in Fig. \ref{flux} is
the expected flux versus nuclearite mass, assuming that the above assumptions
are correct. Obviously, as the mass becomes larger, the production mechanisms for
normal cosmic rays cannot anymore apply to nuclearites.

Different nuclearite flux estimates were recently published \cite{mad}. They are
based on the hypothesis that large nuclearites (with masses $10^{-5} - 10^{-2}$
solar masses) are produced in binary strange stars systems, before their gravitational
collapse. The propagation inside the galaxy considers also the escape, spallation
(through which smaller nuclearites are produced) and  re-acceleration mechanisms.
Nuclearite decays are not considered, as SQM is supposed to be absolutely stable.
The predicted strangelet fluxes around the Earth are presented in Fig. \ref{flux}
for``normal" and CFL strangelets as the dashed and the dot-dashed lines, respectively.
The small differences (that vanish for larger masses, when nuclearites become
completely neutral) originate in the slightly different charge-to-mass ratios (see
Fig. \ref{a-z}).

LMNs would be similar to normal nuclei, except their low Z/A values and, most
likely, different abundances. If they interact with the Earth's atmosphere
in the same way as CR nuclei, they would not reach experiments at mountain altitude.

Two different (and opposite) theoretical scenarios, both consequences of the
hypothesis that SQM is more stable than ordinary nuclear matter,
were introduced in order to allow deep penetration of small nuclearites
 in the atmosphere; none of those mechanisms would allow them anyway to reach
 sea level; such objects could be found only in high altitude experiments.

\subsection{Mass and size decrease of nuclearites during propagation}

In \cite{wilk1} it was assumed that small nuclearites could  penetrate
the atmosphere if  their size and mass are reduced through successive interactions
with the atomic air nuclei.
The proposed scenario is based on the spectator-participant picture.
 Two interaction models are considered: quark-quark (called
``standard"), and collective (called ``tube-like").
 At each
interaction the nuclearite mass is reduced by about the mass of a Nitrogen nucleus
(in the ``standard" model), or by more (in the ``tube-like model"),
while the spectator quarks form a lighter nuclearite that continues its
flight with essentially the same velocity as the initial one.
Once a critical mass is reached  ($A \simeq$ 300 - 400)
neutrons would start to evaporate from strangelets;  for $A < 230$
the SQM would become unstable and decay into normal matter.
In ref. \cite{wilk2} an estimate was made of the
sensitivity of the SLIM experiment \cite{slim}:
the
 mass number of a nuclearite penetrating the atmosphere
down to the Chacaltaya lab would be
one seventh of that it had at the top of the atmosphere.

The CR39 used in SLIM is sensitive to particles with a Restricted Energy
Loss (REL) larger than 200 MeV g$^{-1}$ cm$^2$
\footnote{The threshold of a NTD depends on the etching conditions. A
relatively high threshold for CR39 was
chosen in order to reduce the background
due to recoil tracks, neutron interactions and the ambient radon radioactivity.}.
Table 1 presents
the minimum detectable mass for strangelets reaching SLIM.

\small
\vspace{5mm}
\begin{tabular}{|c|c|c|c|c|}
\hline
A/Z  & $A$ & $A_0$ & $\Phi$ & $\Phi$ \\
hypothesis & (at SLIM) & (top of  &cm$^{-2}$s$^{-1}$sr$^{-1}$ &
 cm$^{-2}$s$^{-1}$sr$^{-1}$\\
  &   & the atmosphere)  & (Ref. \cite{mad}) & (heavy nuclei) \\
\hline
Liquid-drop & & & &\\
or CFL & $\simeq 587$ & $\simeq 4109$ & $6 \times 10^{-12} (CFL)$
 & $2.4 \times 10^{-15}$\\
Refs. \cite{kasuya,madsen} &  &  &$3 \times 10^{-12}$ (normal)  &\\
\hline
``Normal" & & & &\\
nuclearites & $\simeq 210$ & $\simeq 1470$ & $9.5 \times 10^{-12}$ &
$2.8 \times 10^{-12}$ \\
Refs. \cite{heis,madsen} & & & &\\
\hline
\end{tabular}
\begin{center}
\normalsize

Table 1: The minimum strangelet masses
 detectable in SLIM assuming different $A/Z$ relations. The
masses at the top of the atmosphere are estimated as in Ref. \cite{wilk1},
and the expected fluxes are computed as in Ref. \cite{mad}, and assuming the
same mass abundances of ordinary cosmic ray nuclei.
\end{center}


\subsection{Accretion of neutrons and protons during propagation}

A completely different propagation scenario was proposed in \cite{raha}. The
authors assume that small mass nuclearites would pick-up nuclear matter during
interactions with  air nuclei, rather than loosing mass.
After each interaction, the nuclearite mass would increase by about the atomic
mass of Nitrogen, with a corresponding slight reduction of velocity. As the
mass grows larger, the loss in velocity becomes smaller.
They estimate that a
strangelet of an initial $A \simeq 64$ and an electric charge
of about +2 could arrive at about 3600 m a.s.l. with $A \simeq 340$ (3600 m is
the altitude of a proposed
NTD experiment in Sandakphu, India \cite{raha}).
This mechanism would also imply an increase of the electric charge of the strangelet,
 thus an increase of the Coulomb barrier; this may be the main
difficulty of this scenario.
The flux according to \cite{mad} would be of the order of $10^{-9}$
cm$^{-2}$s$^{-1}$sr$^{-1}$, and  higher.

\vspace{3mm}

About 171 m$^2$
 of the SLIM modules exposed for an average time of 3.5 years were removed,
processed and analized. No cadidate survived. The 90\% C.L. flux
upper limit for downgoing nuclearites (LMNs and IMNs) is at the level of $4 \times
10^{-15}$ cm$^{-2}$sr$^{-1}$s$^{-1}$.
This would  disfavor the ``accretion
scenario", and most of the hypotheses quoted in Table 1.

\section{Intermediate mass nuclearites (IMNs)}

In \cite{ruhula} was postulated that elastic collisions
with the atoms and molecules of the traversed medium
 are the only relevant energy loss
mechanism of non-relativistic
nuclearites with large masses,
\begin{equation}
\frac{dE}{dx}=-\sigma \rho v^2,
\label{ruhula1}
\end{equation}
where $\rho$ is the density of the traversed medium, $v$ is the
nuclearite velocity and $\sigma$ is its cross section:
\begin{equation}
\sigma= \left\{ \begin{array}{ll}
       \pi(3M/4 \pi \rho_N)^{2/3}  & \mbox{for $ M \geq 8.4 \times 10^{14}$ GeV
        (corresponding to $R_N \simeq 1$ \AA)} \\
       \pi \times 10^{-16} \mbox{cm}^2 & \mbox{for lower mass nuclearites}
       \end{array}
       \right. ,
\label{ruhula2}
\end{equation}
with $\rho_N = 3.6 \times 10^{16}$ g cm$^{-3}$.  As the
chemical potential of the $s$ quarks in SQM is slightly larger than for
$u$ and $d$ quarks, SQM is always positively charged \cite{ruhula},
thus the cross section for nuclearites with $M < 8.4 \times 10^{14}$ GeV is
determined by their electronic cloud.

The following calculations apply to nuclearites of mass $M$ much larger than typical
nuclear masses and $\beta \simeq 10^{-3}$; effects due to possible
ionization or mass variations
during their flight in the atmosphere are negligible. We also neglect the
gravitational acceleration of nuclearites by the Earth.\footnote{Assuming
a nuclearite
mass of 1 ng (about 5.6$\times 10^{14}$ GeV) arriving at an altitude of
5000 m with $\beta = 10^{-3}$, the gravitational energy gain would
represent less than about $1.5 \times 10^{-3}$ of the energy loss in the atmosphere;
 for $\beta = 10^{-4}$ (near the Makrofol threshold) this ratio
is about 0.15.}

A nuclearite of mass $M$
entering  the atmosphere with an initial
velocity $v_0 << c$, after crossing a depth $L$ will be slowed down to
\begin{equation}
v(L) = v_0 e^{-\frac{\sigma}{M}\int_0^L{\rho dx}}
\label{vit}
\end{equation}
where $\rho$ is the air density at different depths, and $\sigma$ is
the interaction cross section of Eq. \ref{ruhula2}.

We consider the parametrization of the standard atmosphere from \cite{shibata}:
\begin{equation}
\rho(h) = a e^{-\frac{h}{b}} = a e^{-\frac{H-L}{b}},
\end{equation}
where the constants are
$a = 1.2 \times 10^{-3}$g cm$^{-3}$ and $b \simeq 8.57 \times 10^5$ cm; $H$
is the total
height of the atmosphere ( $\simeq$ 50 km).
The integral in Eq. 3 may be solved analytically:
\begin{equation}
\int_0^L{\rho dx} = a b e^{-\frac{H}{b}} \left(e^{\frac{H-h}{b}} -1 \right).
\end{equation}

\begin{figure}[htb]
\vspace{-35mm}
\begin{center}
\includegraphics[angle=0, width=1\textwidth]{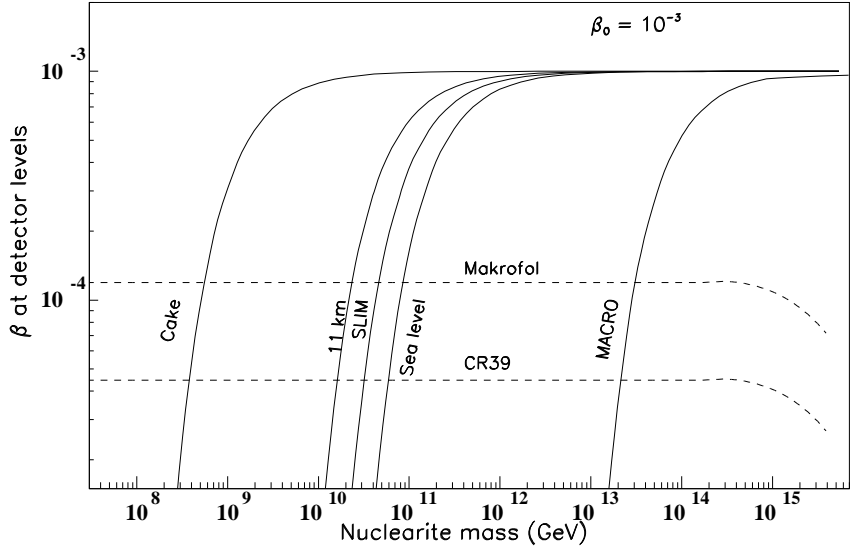}
\caption{Solid lines: arrival velocities of IMNs at different depths
versus nuclearite mass, assuming an
initial velocity outside the atmosphere of $\beta = 10^{-3}$.
The nuclearites are supposed to come from above, close to the vertical direction.
The dashed lines
show the detection thresholds in CR39 (in the SLIM etching conditions) and Makrofol.}
\label{betam}
\end{center}
\end{figure}

Fig. \ref{betam}
 shows the velocity with which nuclearites of different masses reach heights
 corresponding to typical balloon experiments
 (for instance CAKE, 40 km \cite{cake}),
 possible experiments using
civilian airplanes (11 km), the Chacaltaya lab (SLIM, 5.29 km \cite{slim})
and at sea level. A
computation valid for  MACRO \cite{macro} (at a depth of 3400 mwe) is also included.
The velocity thresholds for detection
in CR39 (corresponding to REL = 200 MeV g$^{-1}$ cm$^2$)\footnote{We
 recall that in the low background conditions of the Gran Sasso Lab,
in the case of the MACRO experiment the CR39 detection threshold was set to
50  MeV g$^{-1}$ cm$^2$}
 and in Makrofol
(REL = 2500 MeV g$^{-1}$ cm$^2$) are shown as the dashed curves.

The decrease of the velocity thresholds for nuclearite
masses larger than $8.4 \times
10^{14}$ GeV is due to the
change in the nuclearite cross section, according  Eq.
\ref{ruhula2}.

An experiment at the Chacaltaya altitude lowers the minimum detectable
nuclearite mass by a factor of about 2 with respect to an experiment
performed at sea level.
If the mass abundance of nuclearites decreases strongly with increasing
mass
this could yield an
important increase in sensitivity.

\begin{figure}[htb]
\vspace{-35mm}
\begin{center}
\includegraphics[angle=0, width=1\textwidth]{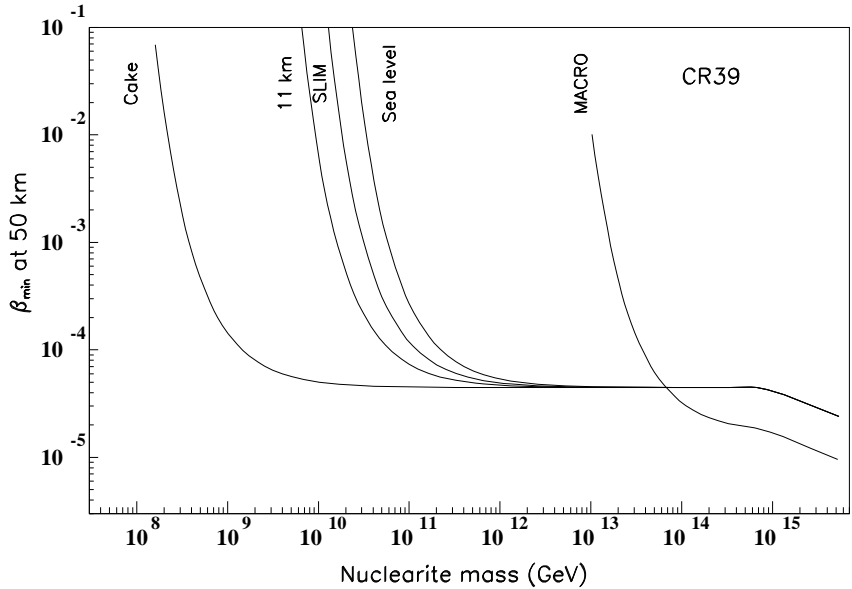}
\caption{Nuclearite detection conditions in CR39, for experiments located
at different altitudes.}
\label{acces1}
\end{center}
\end{figure}

The nuclearite detection conditions in CR39
(expressed as the minimum entry velocity at the top of the atmosphere
versus the nuclearite mass) for different experimental
locations is shown in Fig. \ref{acces1}. In this case, the constraint
 is that nuclearites have the minimum velocity at the
detector level in order to produce a track; we remind that for all experiments
 the REL threshold for detection in CR39 is set to 200 MeV g$^{-1}$ cm$^2$
(it was 50 MeV g$^{-1}$ cm$^2$ in the case of MACRO).

\begin{figure}[htb]
\vspace{-15mm}
\begin{center}
\includegraphics[angle=0, width=1\textwidth]{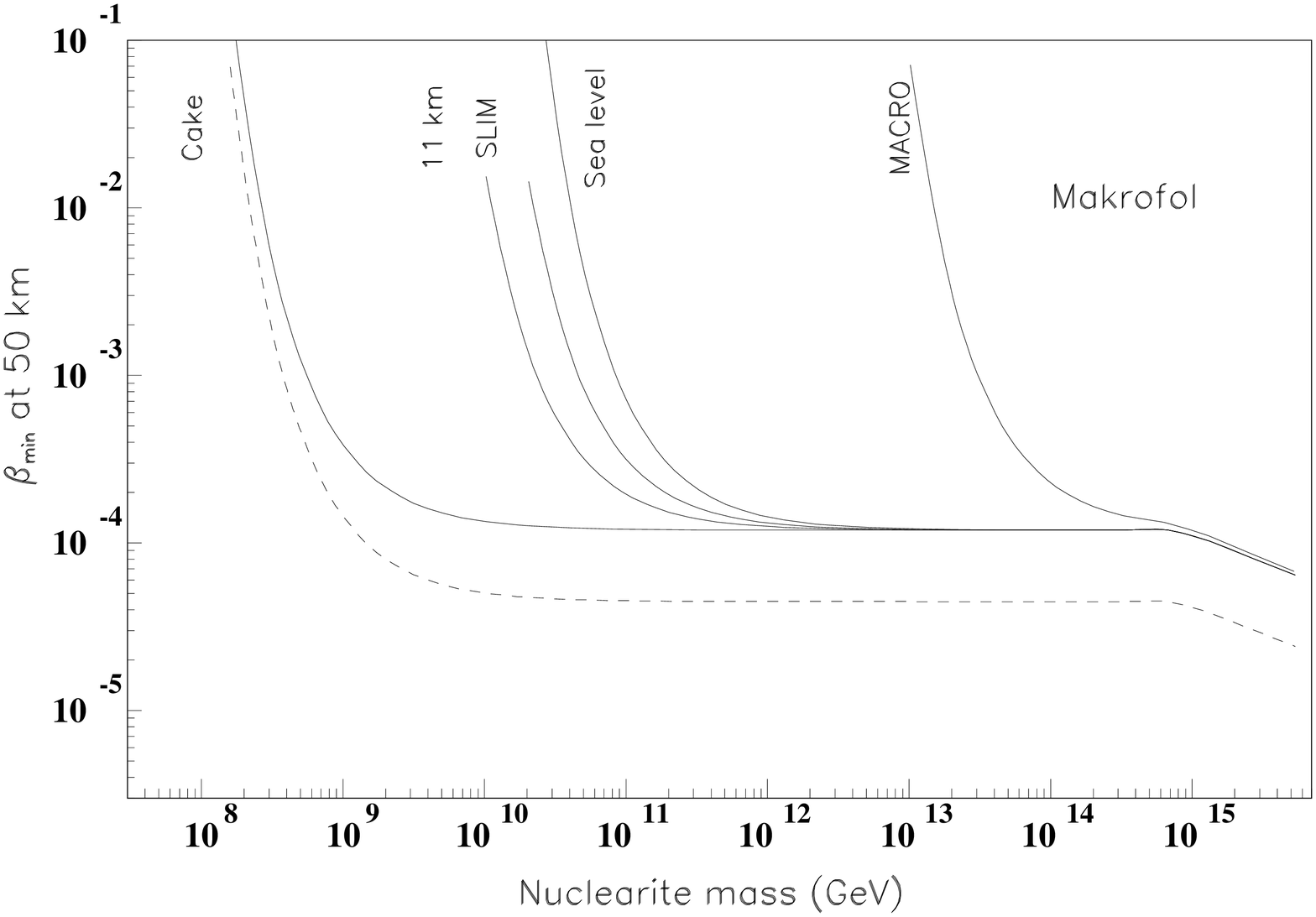}
\caption{Nuclearite detection conditions in Makrofol, for experiments located
at different altitudes. The dashed curve (shown for comparison) corresponds to
CR39 in CAKE.}
\label{acces2}
\end{center}
\end{figure}

Fig. \ref{acces2} shows the same for the Makrofol track
etch detector. The detection condition corresponding to CR39 at balloon altitude
(CAKE) is also shown.

\section{Conclusions}

SLIM is a large area NTD experiment, taking data at the Chacaltaya
Cosmic Ray Laboratory. In this note we investigated the possibility to search
for nuclearites with SLIM, assuming two nuclearite mass regions.

{\bf Low mass nuclearites} could reach mountain altitudes assuming some
peculiar interaction mechanisms in the atmosphere. They would produce in NTDs tracks
similar to those expected from  fast monopoles or
relativistic heavy nuclei\footnote{Note that relativistic nuclei present in
the CR cannot penetrate the Earth's atmosfere till the Chacaltaya level.} \cite{slim}.
SLIM will reach a sensitivity at the level of about $10^{-15}$
cm$^{-2}$s$^{-1}$sr$^{-1}$ for a flux of nuclearites coming from above. In the
absence of a LMN candidate, SLIM would rule out the propagation
mechanisms hypothesized.
SLIM will be also sensitive to different strangelet structure hypotheses (``normal"
or CFL, different $Z/A$ predictions), and could validate or not the production
and propagation model proposed in \cite{mad}.

{\bf Intermediate mass nuclearites}, entering the Earth atmosphere with typical
galactic velocities  might be detected  by large
area  NTDs as SLIM.
The minimum detectable nuclearite mass is very sensitive to the experiment location:
 an underground experiment like MACRO could search for nuclearites with
$M \geq 10^{14}$ GeV, the minimum mass accessible to ANTARES \cite{antares}
is of few $10^{13}$
GeV. Detectors at ground level, or better at mountain altitudes like SLIM,
would decrease the mass threshold to few $10^{10}$ GeV; balloon experiments are
needed to reach few $10^8$ GeV, while lower mass searches have to be
done outside the Earth atmosphere.

SLIM
is sensitive to non relativistic
($\beta \lesssim 10^{-3}$) IMNs with masses larger than $3 \times 10^{10}$
GeV; the large REL of IMNs in NTDs and their property to produce identical tracks
in all the detector sheets in a stack could yield experimental signatures with  low
background.

Nuclearites with masses between few TeV and about $10^8$ GeV are not
considered at this time. In the low part of this mass range, they could still
be accelerated to relativistic velocities by  cosmic fields, and could be
detected as LMNs in high altitude experiments. For larger nuclearite masses,
one would expect the velocity to get close to the galactic one, and thus such
nuclearites would not be able to reach the detector. A search for
nuclearites in this mass range could, in principle, be done in space experiments, such
as the AMS detector on board of the International Space Station \cite{ams}.

\begin{figure}[htb]
\vspace{-15mm}
\begin{center}
\includegraphics[angle=0, width=1.1\textwidth]{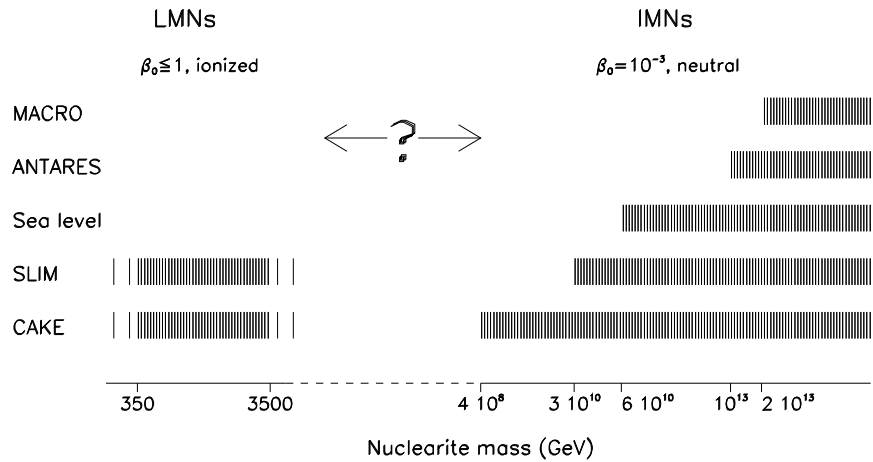}
\caption{Approximate nuclearite mass regions, accessible to different experiments,
(The drawing is not to scale).}
\label{conc}
\end{center}
\end{figure}

Fig. \ref{conc}  summarizes the conclusion, showing the accessible nuclearite
mass regions for different experiments.

\end{document}